\documentclass[fleqn]{article}

\usepackage{amsmath,amssymb}
\usepackage{graphicx}

\def\tlabel#1{\label{#1}}
\def\bref#1#2#3{{#1}~(#2)~#3}
\def\tref#1{(\ref{#1})}

\begin{document}

% Titlepage and abstract
\title{{\sf  
Constraints on Higher-Order Perturbative Corrections in $b\to u$ Semileptonic Decays
 from  Residual Renormalization-Scale Dependence}}
\author{M.R.~Ahmady$^1$, F.A.~Chishtie$^2$, V.~Elias$^2$, A.H.~Fariborz$^3$, \\
D.G.C.~McKeon$^2$, T.N.~Sherry$^4$, 
T.G.~Steele$^5$}
\footnotetext[1]{Department of Physics, Mount Allison University, Sackville, NB~~E4L 1E6, Canada}
\footnotetext[2]{Department of Applied Mathematics,
The University of Western Ontario,
London, ON~~ N6A 5B7, Canada}
\footnotetext[3]{Dept.\ of Mathematics/Science, State Univ.\ of New York Institute of Technology, Utica, NY~~13504-3050, USA}
\footnotetext[4]{Department of Mathematical Physics, National University of Ireland, Galway, Ireland }
\footnotetext[5]{Department of Physics \& Engineering Physics,
University of Saskatchewan,
Saskatoon, SK~~ S7N 5E2, Canada}
 
\maketitle

\begin{abstract}
The constraint of a progressive decrease in residual renormalization scale dependence with 
increasing loop order is developed as a method for obtaining bounds on unknown 
higher-order perturbative corrections 
to renormalization-group invariant quantities.
This technique is applied to the inclusive semileptonic process  
$b\rightarrow  u \bar\nu_\ell\ell^-$ (explicitly known to two-loop order)  
in order to obtain bounds on  the three- and four-loop perturbative contributions
that are not accessible via the renormalization group.  Combining this technique with  the principle of 
minimal sensitivity, we obtain an estimate  for the perturbative contributions to   
$\Gamma\left(b\rightarrow  u \bar\nu_\ell\ell^-\right)$ that incorporates theoretical uncertainty 
from as-yet-undetermined higher order QCD corrections.
\end{abstract}

\noindent
A variety of techniques have been developed for obtaining estimates of higher-loop  corrections in 
perturbative QCD, including the principle of minimal sensitivity (PMS) \cite{PMS}, the fastest apparent convergence (or effective charges) approach \cite{FAC}, nonabelianization methods 
\cite{Maxwell},  
 Pad\'e approximations \cite{Pade}, and  renormalization-group supplemented Pad\'e approximants \cite{Elias}.  Of particular relevance is the PMS ansatz which proposes that 
renormalization-scheme independence, which at lower-loop levels translates into renormalization-scale ($\mu$) independence, provides the optimum perturbative prediction of a QCD observable.  

In a particular renormalization scheme (such as $\overline{{\rm MS}}$), the minimal sensitivity principle identifies the appropriate 
choice of renormalization scale $\mu$ for a physical observable as the  value at which the observable is independent of $\mu$, providing a method for dealing with  the residual renormalization scale dependence that exists in a
perturbative calculation truncated to any given order.  In explicit calculations, such residual scale
dependence  decreases as higher-order corrections are included; if
 a renormalization-group-invariant quantity $\Gamma$ is truncated   at order $\left(\alpha/\pi\right)^n$, 
one can easily demonstrate that $\mathrm{d}\Gamma/\mathrm{d}\mu$ is order $\left(\alpha/\pi\right)^{n+1}$. 
Consequently,  
perturbative predictions become virtually $\mu$-independent 
at sufficiently high orders.   For example, in inclusive semileptonic $b\to u$ decays, a clear progressive flattening of the decay rate as a function of $\mu$ is 
observed in the explicit calculations up to two-loop order \cite{VanRit}, and this property persists when 
a Pad\'e estimate of the three-loop correction is included \cite{Ahmady}.

In this paper we demonstrate for semileptonic $b\to u$ decays that the  progressive 
decrease in  renormalization-scale dependence with increasing loop-order  
places  bounds on unknown higher-order perturbative coefficients.  
Combined  with  minimal-sensitivity, the prediction of the decay rate devolving 
from these bounds leads to an estimate of higher-order perturbative corrections to the 
decay rate. 
 
The perturbative contributions to the
 inclusive semileptonic $b\rightarrow   u$ decay rate
$\Gamma(b\rightarrow u\bar\nu_\ell \ell^-)$
 may be expressed as \cite{VanRit}
\begin{equation}
\begin{split}
\frac{1}{K}\Gamma\left(\mu, m_b(\mu),x(\mu)\right)=m_b^5(\mu)
\biggl(1+&\left[a_0-a_1\log(w)\right] x(\mu)+
\left[ b_0-b_1\log(w)+b_2\log^2(w)\right] x^2(\mu)
\\
 +&\left[ c_0-c_1\log(w)+c_2\log^2(w)-c_3\log^3(w)\right] x^3(\mu)
\\
+&\left[ d_0-d_1\log(w)+d_2\log^2(w)-d_3\log^3(w)+d_4\log^4(w)\right] x^4(\mu)
+\ldots \biggr)
\end{split}
\label{basic_gamma}
\end{equation}
where
\begin{equation}
x(\mu)\equiv \frac{\alpha_s(\mu)}{\pi}\quad , \quad
w=w(\mu,m_b(\mu))\equiv \frac{m_b^2(\mu)}{\mu^2}\quad ,\quad  
K\equiv \frac{G_F^2\left| V_{ub}\right|^2}{192\pi^3}\quad .
\label{gamma_defns}
\end{equation}
In the 
 $\overline{{\rm MS}}$ scheme in which $m_b$ is identified with a 
scale-dependent (running) $b$-quark mass for four or five active flavours, the one- and 
two-loop order 
 coefficients within 
\tref{basic_gamma} are  given by \cite{VanRit}
\begin{gather}
{\rm all}~ n_f:\quad a_0=4.25360\quad ,\quad a_1=5\quad , \nonumber\\
 n_f=5:\quad b_0=26.7848\quad ,\quad b_1=36.9902\quad ,\quad b_2=17.2917\quad ,\nonumber \\
 n_f=4: \quad b_0=25.7547\quad ,\quad b_1=38.3935\quad ,\quad b_2=17.7083\quad .
\label{pert_coeffs}
\end{gather}
The running mass $m_b(\mu)$ appearing in \tref{basic_gamma} is also known to full four-loop order 
\cite{anomdim}:
\begin{gather}
m_b\left[x(\mu)\right] = m_b\left[x\left(\mu_0\right)\right] 
\frac{c\left[x(\mu)\right]}{c\left[x\left(\mu_0\right)\right]}\quad,
\\
n_f = 4: \quad c[x]  =  x^{12/25} \left[ 1 + 1.01413x + 1.38920 x^2
 +   1.09052 x^3 +\ldots \right] ,
\\
n_f = 5: \quad c[x]  =  x^{12/23} \left[ 1 + 1.17549x + 1.50071 x^2
 +   0.172486 x^3 +\ldots \right] \quad .
\end{gather}

Renormalization-group invariance of the decay rate $\Gamma$ determines a subset of the unknown three-loop
$c_k$ coefficients and four-loop $d_k$ coefficients.  From the renormalization-group equation
\begin{equation}
\begin{split}
0&=\mu^2\frac{\mathrm{d}}{\mathrm{d}\mu^2}\Gamma\left(\mu,m_b(\mu),x(\mu)\right)
\\
&=\left[\mu^2\frac{\partial}{\partial\mu^2}-\left(\gamma_0 x+\gamma_1x^2+\gamma_2 x^3+\ldots \right)
m_b\frac{\partial}{\partial m_b}-\left(\beta_0x^2+\beta_1 x^3+\ldots\right)\frac{\partial}{\partial x}\right] \Gamma\quad ,
\end{split}
\label{RG}
\end{equation}
we find the following expressions for  terms proportional to $x^3$ and $x^4$ in the final line of 
\tref{RG}:
\begin{gather}
c_1=2b_0\beta_0+a_0\beta_1+\gamma_0\left(5b_0-2b_1\right)+\gamma_1\left(5a_0-2a_1\right)+5\gamma_2\quad ,
\label{RGc1} \\
c_2=\frac{1}{2}\left[2b_1\beta_0+a_1\left(\beta_1+5\gamma_1\right)+\gamma_0\left( 5 b_1-4 b_2\right)\right] \quad ,
\label{RGc2}\\
c_3=\frac{b_2}{3}\left(2\beta_0+5\gamma_0\right)\quad,
\label{RGc3}\\
d_1=\beta_2a_0+2\beta_1b_0+3\beta_0c_0+\gamma_0\left(5c_0-2c_1\right)
+\gamma_1\left(5b_0-2b_1\right)+\gamma_2\left(5a_0-2a_1\right)+5\gamma_3
\label{RGd1}\\
d_2=\frac{1}{2}\beta_2a_1+\beta_1b_1+\frac{3}{2}\beta_0c_1+\gamma_0\left(\frac{5}{2}c_1-2c_2\right)
+\gamma_1\left(\frac{5}{2}b_1-2b_2\right)+\frac{5}{2}\gamma_2a_1
\label{RGd2}\\
d_3=\beta_0c_2+\frac{2}{3}\beta_1b_2+\gamma_0\left(\frac{5}{3}c_2-2c_3\right)+\frac{5}{3}\gamma_1b_2
\label{RGd3}\\
d_4=\frac{3}{4}\beta_0c_3+\frac{5}{4}\gamma_0c_3
\label{RGd4}
\end{gather}
Upon substitution of  \tref{pert_coeffs} and 
 the four-loop 
 $\overline{{\rm MS}}$  results for the $\beta$ function \cite{beta} and anomalous mass dimension
$\gamma$ 
\cite{anomdim} into (\ref{RGc1}--\ref{RGd4}), we obtain the following  numerical values for these higher-loop coefficients:
\begin{gather}
 c_1^{(4)} = 263.839~,~ c_2^{(4)} = 194.234~,~ c_3^{(4)} = 54.1087
\label{c4f}\\
 d_1^{(4)}=11.25c_0^{(4)}+103.081~,~d_2^{(4)}=1580.26~,~d_3^{(4)}=765.844~,~d_4^{(4)}=152.181
\label{d4f}\\
 c_1^{(5)} = 249.592~,\quad c_2^{(5)} = 178.755~, \quad c_3^{(5)} = 50.9145
\label{c5f}\\
 d_1^{(5)}=10.75c_0^{(5)}-8.28683~,~d_2^{(5)}=1376.68~,~d_3^{(5)}=667.838~,~d_4^{(5)}=136.833
\label{d5f}
\end{gather} 
where the superscript denotes the number of active flavours (either $n_f=4$ or $n_f=5$).  
In the energy region spanning the threshold between $n_f=4$  and $n_f=5$ there are four higher-loop parameters that remain undetermined: $\left\{ c_0^{(4)},~ c_0^{(5)}, ~ d_0^{(4)},~ d_0^{(5)}\right\}$.
However,  continuity of the decay rate at this threshold in conjunction with the 
coupling-constant and quark-mass 
threshold matching conditions imposed at $\mu=m_b$ \cite{matching},\footnote{In the $\overline{MS}$
scheme the scale $\mu=m_b$ is defined by $m_b(m_b)=m_b$.}
\begin{gather}
x^{(4)}\left(m_b\right)=x^{(5)}\left(m_b\right)\left[1+0.1528\left[x^{(5)}\left(m_b\right)\right]^2
+0.633\left[x^{(5)}\left(m_b\right)\right]^3\right] 
\label{match_alpha}\\
m_b^{(4)}\left(m_b\right)=m_b^{(5)}\left(m_b\right)\left[1+0.2060\left[x^{(5)}\left(m_b\right)\right]^2+
1.9464\left[x^{(5)}\left(m_b\right)\right]^3\right] \quad ,
\label{match_mb}
\end{gather}
effectively reduces this set to two unknown parameters.
 Using the benchmark value $\alpha_s\left(M_z\right)=0.119$ \cite{pdg}, $m_b=4.2\,{\rm GeV}$ \cite{bmass}, and the four-loop $\beta$ function, we first find that
\begin{gather}
x^{(5)}\left(4.2\,{\rm GeV}\right)=0.07261 \quad ,\quad x^{(4)}\left(4.2\,{\rm GeV}\right)=0.07269\\
m_b^{(4)}\left(4.2\,{\rm GeV}\right)=4.208\,{\rm GeV}\quad .
\end{gather}
Imposing continuity of the decay rates to ${\cal O}\left(x^3\right)$ at this flavour threshold then yields 
\begin{equation}
c_0^{(5)}=1.012c_0^{(4)}+15.66\quad .
\tlabel{c05c04}
\end{equation}
Such a continuity condition cannot be extended to ${\cal O}\left(x^4\right)$, because 
 the first
unknown term in the mass threshold condition \tref{match_mb} contributes at 
${\cal O}\left(x^4\right)$ to the
 $d_0$ coefficient.  In the absence of this information, we 
assume 
\begin{equation}
d_0^{(4)}\approx d_0^{(5)}\quad , 
\end{equation}
reflecting the near-equivalence of the known coefficients $\left\{d_2,~d_3,~d_4\right\}$ 
for four and five active flavours.\footnote{In \protect\tref{match_mb}, if the coefficient of 
$\left[x^{(5)}\left(m_b\right)\right]^4$ were $20$, reflecting a factor of ten increase between the two previous
orders, the difference between $d_0^{(4)}$ and $d_0^{(5)}$ would be approximately $150$, which is small 
compared to the scales of $d_0$ we obtain in our analysis below. 
}
Thus the RG analysis combined with continuity of the decay rate at flavour thresholds implies that 
the four-loop expression \tref{basic_gamma} effectively has only  two undetermined parameters 
$\left\{c^{(4)}_0,~d^{(4)}_0\right\}$
in the energy range spanning the threshold between $n_f=4$ and $n_f=5$.

The decay rate $\Gamma$ is a truncated perturbation series which necessarily exhibits residual renormalization scale dependence.  In general if $\Gamma$ is known to ${\cal O}\left(x^n\right)$, then 
$\mathrm{d}\Gamma/\mathrm{d}\mu={\cal O}\left(x^{n+1}\right)$. Consequently, for any reasonable perturbation series,  the residual scale 
dependence diminishes with increasing loop order.  A measure of the residual scale dependence in the natural energy region\footnote{This energy region is exactly that considered
in \protect\cite{VanRit} where the residual scale dependence of the two-loop decay rate was originally  considered.} 
 $m_b/2<\mu<2m_b$
 is provided by the deviation of $\Gamma$ from its average  value
\begin{equation}
\chi^2=\frac{1}{\frac{3}{2}m_b}\int\limits_{m_b/2}^{2m_b}\left(\frac{\Gamma(\mu)}{\left\langle\Gamma\right\rangle}-1\right)^2
\,\mathrm{d}\mu \quad ,
\tlabel{chi2}
\end{equation}
where the pre-factor of $3m_b/2$ leads to a dimensionless $\chi^2$, and 
where $\left\langle\Gamma\right\rangle$ is the average value of $\Gamma$ over the same  energy interval as \tref{chi2}:
\begin{equation}
\left\langle\Gamma\right\rangle=\frac{1}{\frac{3}{2}m_b}\int\limits_{m_b/2}^{2m_b}\Gamma(\mu)
\,\mathrm{d}\mu  \quad .
\tlabel{GammaBar}
\end{equation}
The progressive decrease in $\Gamma$'s scale dependence implies that   $\chi^2$
must decrease as the loop order of $\Gamma$ is increased.  However, since at three-loop 
order $\Gamma$ (and hence $\chi^2$) will depend on the parameter $c_0^{(4)}$, and at four-loop order will depend on   $\left\{c_0^{(4)},~d_0^{(4)}\right\}$, the  progressive decrease in 
$\chi^2$ as loop-order increases necessarily provides constraints on these unknown higher-loop parameters.

\begin{figure}[hbt]
\centering
\includegraphics[scale=0.7]{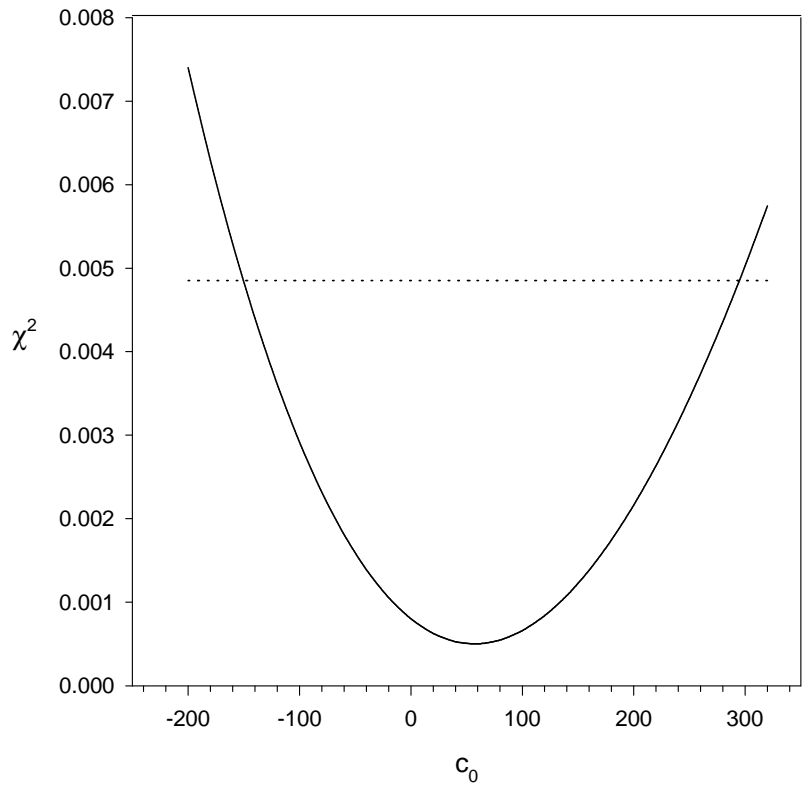}
\caption{The quantity $\chi^2$  as a function of $c_0^{(4)}$ (solid curve) for the three-loop 
contributions 
to $\Gamma$.  The straight dashed line represents the two-loop $\chi^2$, and the criterion that the three-loop $\chi^2$ is smaller than the  two-loop $\chi^2$ constrains $c_0^{(4)}$ to the region between the intersection points of the three- and two-loop curves.
}
\label{c0fig}
\end{figure}

To obtain such constraints, we use
 the central values
 $\alpha_s\left(M_z\right)=0.119$ \cite{pdg}, $m_b=4.2\,{\rm GeV}$ \cite{bmass}, 
the four-loop $\beta$ and $\gamma$ functions, and the threshold matching conditions
\tref{match_alpha}, \tref{match_mb} to evaluate $\alpha(\mu)$ and $m_b(\mu)$. We then utilize  
 discretization to evaluate the integrals  \tref{chi2} and \tref{GammaBar} and obtain
$\chi^2$ values shown in Figure \ref{c0fig}.  The two-loop result is by definition 
independent of 
 $c_0^{(4)}$;  the three-loop result depends on $c_0^{(4)}$ as displayed in Figure \ref{c0fig}.  
The requirement  that the 
three-loop $\chi^2$ is less than the two-loop $\chi^2$ is satisfied provided 
\begin{equation}
-150<c_0^{(4)}<290\quad .
\tlabel{c04bound}
\end{equation}
The Pad\'e estimate $c_0^{(4)}=188$ of \cite{Ahmady} is well within  this interval.
Combining the result \tref{c04bound}  with \tref{c05c04} we find the following constraint on  $c_0^{(5)}$:
\begin{equation}
-136<c_0^{(5)}<309\quad ,
\end{equation}
which, in the notation  $c_0^{(5)}=200+\Delta$,  corresponds to  $-336<\Delta<109$ 
that includes the range of  
 estimates  $-100\le \Delta\lesssim 80$ discussed in  \cite{Melnikov}. The optimal value of $c_0^{(4)}$, {\it i.e.},
the value that minimizes $\chi^2$, occurs at  $c_0^{(4)}=57$  ($c_0^{(5)}=74$) corresponding 
most closely to the large-$\beta_0$ estimate ($\Delta=-100$) of \cite{Melnikov,Beneke}.

\begin{figure}[hbt]
\centering
\includegraphics[scale=0.7]{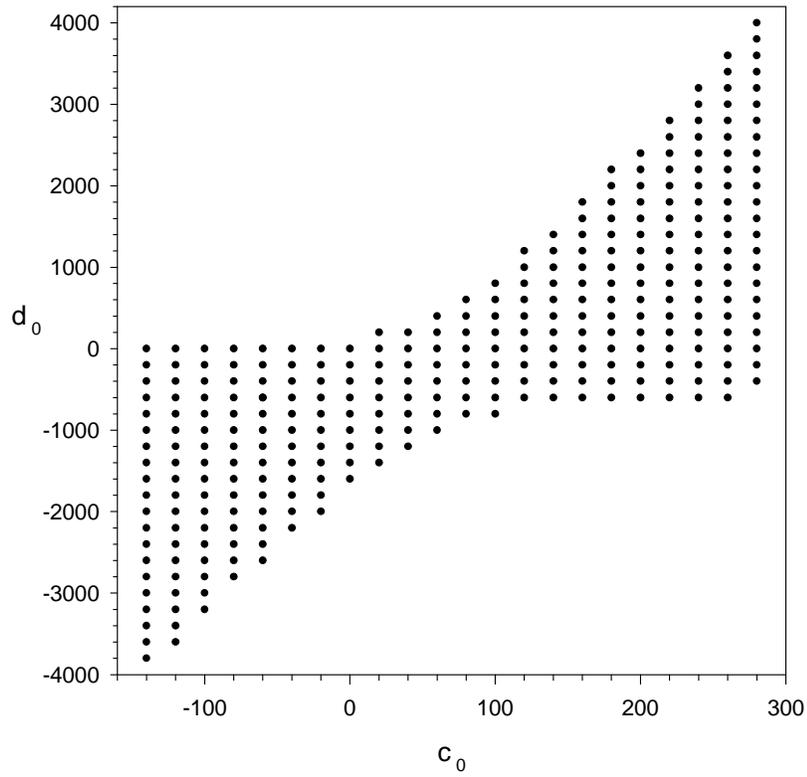}
\caption{The dots correspond to points within  
$\left\{c_0^{(4)},~d_0^{(4)}\right\}$ parameter space 
for  which $\chi^2$ decreases  as loop-order increases from two to four.  
The sharp cutoff in the $c_0^{(4)}$ direction
corresponds to the range obtained from Figure \protect\ref{c0fig}.
}
\label{c0d0fig}
\end{figure}

\begin{figure}[hbt]
\centering
\includegraphics[scale=0.7]{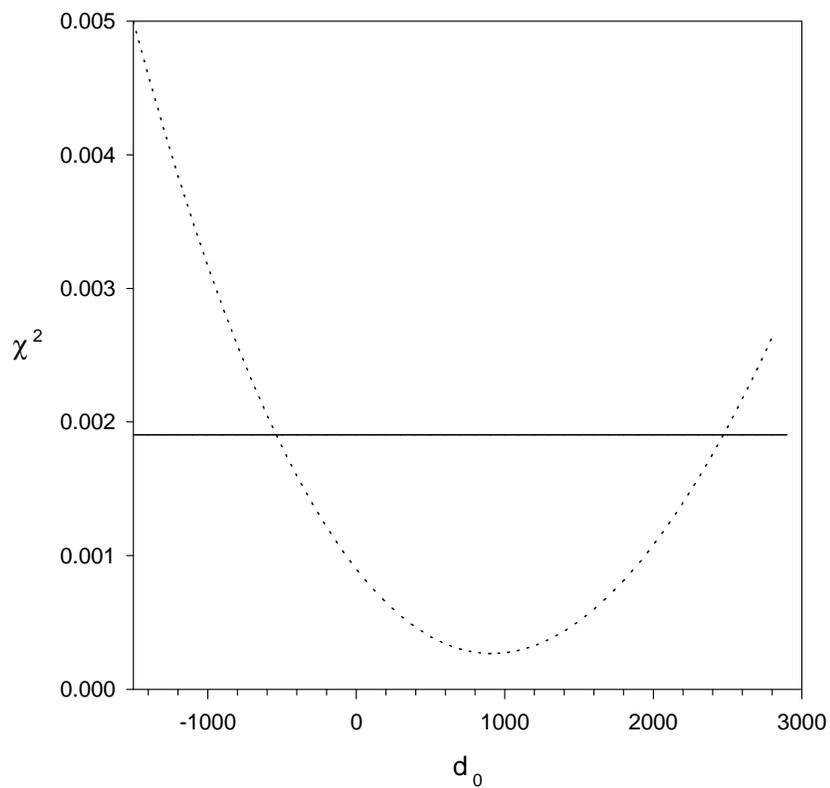}
\caption{
The quantity $\chi^2$  as a function of $d_0^{(4)}$ (dashed curve) for the four-loop 
contributions 
to $\Gamma$ after input of the Pad\'e estimate $c_0^{(4)}=188$ .  
The straight solid line represents the three-loop $\chi^2$. The criterion that the 
four-loop $\chi^2$ is smaller  three-loop $\chi^2$  constrains $d_0^{(4)}$ to the 
region between the intersection points of the curves.
}
\label{d0fig}
\end{figure}

The progressive decrease in $\chi^2$ to four-loop order restricts
$\left\{c_0^{(4)},~d_0^{(4)}\right\}$ parameter space to the region indicated in Figure \ref{c0d0fig}.  
Further insight into the allowed $d_0^{(4)}$ range can be obtained by employing  the Pad\'e
estimate $c_0^{(4)}=188$ \cite{Ahmady}  
to determine a constant 
 three-loop value for $\chi^2$. Upon
comparison  with the four-loop $\chi^2$ dependence on the unknown parameter  $d_0^{(4)}$, 
we find from Figure \ref{d0fig} that $\chi^2$ continues to decrease with increasing order provided
\begin{equation}
-500\lesssim d_0^{(4)}\lesssim 2500\quad .
\tlabel{d04bound}
\end{equation}

\begin{figure}[hbt]
\centering
\includegraphics[scale=0.7]{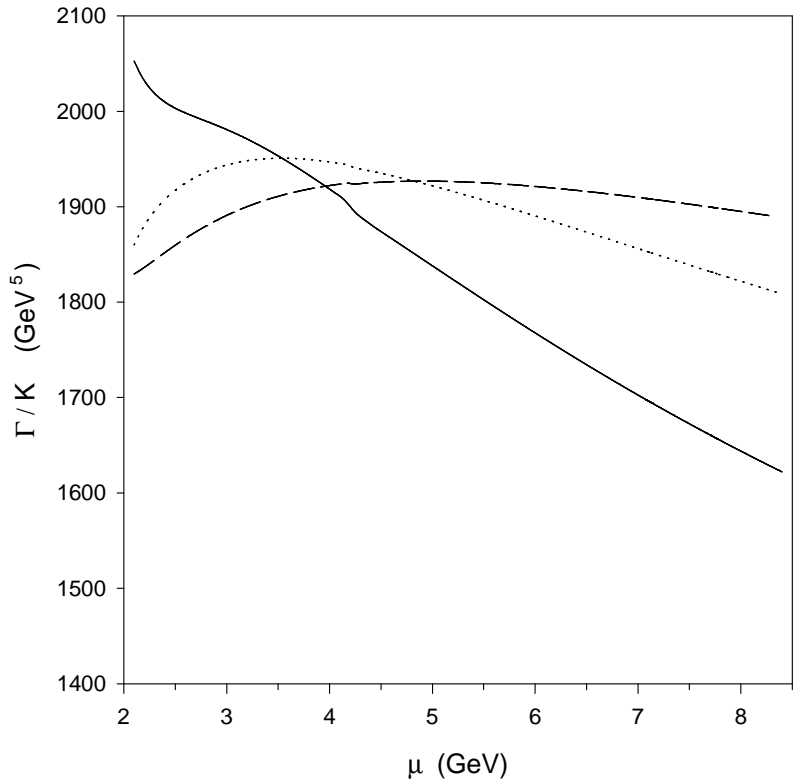}
\caption{
Two-loop (solid curve), three-loop (dotted curve) and four-loop (dashed-curve) values for 
the reduced rate $\Gamma/K$  with  $c_0^{(4)}=80$  
(corresponding to a central value $\Delta=114$ from 
\protect\cite{Melnikov}) and $d_0^{(4)}=-500$ (central value  in the 
Fig.\ \protect\ref{c0d0fig}  parameter space for the chosen $c_0^{(4)}$).  
Progressive decrease of residual scale dependence with increasing loop order is  evident, as is the
existence of PMS points for the three- and four-loop curves.  Note the absence  of a PMS point for the two-loop curve.
}
\label{rate_fig}
\end{figure}

Although the range of parameter space 
in Figure \ref{c0d0fig} [and in \tref{c04bound}--\tref{d04bound}]
might superficially be dismissed as too large to be of interest, it actually 
provides valuable control over higher-order corrections
in the prediction of the
actual phenomenological  decay rate.  If we extract the 
minimal-sensitivity  prediction of the decay rate
over the entire parameter space of Figure \ref{c0d0fig}, 
\footnote{Minimal sensitivity occurs when $\mathrm{d}\Gamma/\mathrm{d}\mu=0$
 in the region $m_b/2\le \mu\le 2m_b$.
When there exists more 
than one critical point, we choose the one with the smallest second derivative, 
corresponding to the least sensitive choice.}
we obtain the following range for the perturbative contributions to the inclusive semileptonic decay rate:
\begin{equation}
\frac{\Gamma}{K}=2050\pm 270\,{\rm GeV^5}\quad .
\tlabel{final_value}
\end{equation}
By contrast, the perturbative two-loop rate, obtained by truncation of the series
\tref{basic_gamma} after its (fully known) ${\cal O}\left(x^2\right)$ contributions, does not exhibit
minimal sensitivity at all in the $m_b/2\le \mu \le 2 m_b$ region, but is seen to decrease with increasing 
$\mu$ from $2053\,{\rm GeV^5}$ to $1622\,{\rm GeV^5}$. Indeed, one unanticipated effect of including 
three- and four-loop coefficients, both known $\left\{c_{1\mbox{--}3},~d_{2\mbox{--}4}\right\}$
and unknown, is to ensure that minimal sensitivity occurs in the range $m_b/2\le \mu\le 2m_b$ over almost 
all of the parameter space of Figure \ref{c0d0fig}.  This property is shown in Figure 
\ref{rate_fig} which compares the two- three- and four-loop contributions to the reduced rate for
a typical point in the Fig.\ \ref{c0d0fig} parameter space.  

The  central value of the bound \tref{final_value} 
corroborates the  Pad\'e estimate $2071\,{\rm GeV^5}$ obtained in \cite{Ahmady}.
The $\pm 13\%$ uncertainty in the rate \tref{final_value}, obtained from the combined constraints of a progressive decrease in residual scale dependence and minimal sensitivity,  
is a genuine reflection of the uncertainty following from unknown higher-order QCD corrections. 
Such corrections, which typically are legislated away
via series truncation after known-order contributions, cannot be
disregarded as a source of additional uncertainty in predictions following
from perturbative quantum field theory.

\smallskip
\noindent
{\bf Acknowledgments:}  We are grateful for the hospitality of the KEK Theory Group, where this research 
was initiated with financial  support
from 
the  International Opportunity Fund of the
 Natural Sciences and Engineering Research Council of Canada (NSERC) 
and the International Collaboration Programme
2001 of Enterprise Ireland.

\clearpage

\clearpage

\clearpage

\end{document}